\documentclass[12pt]{article}
\usepackage{amssymb}
\usepackage{amsfonts}
\usepackage{youngtab}
\usepackage{amsmath,amssymb,latexsym,cite}
\usepackage{amsthm}
\usepackage{cite}
\usepackage[a-1b]{pdfx}    
\usepackage[]{hyperref}
\input xy
\xyoption{all}

\numberwithin{equation}{section}

\begin{document}

\vspace*{0.5in}

\begin{center}

{\large\bf Notes on noninvertible quantum symmetries in two dimensions}

P.~Banerjee, E.~Sharpe

\begin{tabular}{l}
Department of Physics MC 0435 \\
850 West Campus Drive \\
Virginia Tech \\
Blacksburg, VA 24061 
\end{tabular}

{\tt pinakb24@vt.edu},
{\tt ersharpe@vt.edu}

\end{center}

In this note we describe a systematic procedure for computing partition functions of noninvertible Rep ($G$) quantum symmetry actions on two-dimensional
nonabelian $G$ orbifolds.  We apply this procedure to multiplicity-free examples for several nonabelian groups, and check explicitly in those examples that the partition function of a two-dimensional
Rep ($G$)-gauged $G$ orbifold, for nonabelian $G$, matches that of the original theory, as expected on general principles.  This fills a minor gap in the literature, making quantum symmetries in two-dimensional nonabelian orbifolds more concrete.

\begin{flushleft}
    June 2026
\end{flushleft}

\newpage

\tableofcontents

\newpage

\section{Introduction}

Quantum symmetries in two-dimensional abelian orbifolds have long been known, see for example \cite{Vafa:1989ih},
\cite[section 8.5]{Ginsparg}.  These are symmetries that multiply twist fields of a $G$ orbifold by phases.  Orbifolding the orbifold by the quantum symmetry returns the original theory. Somewhat more recently, it has been argued that quantum symmetries also exist in two-dimensional nonabelian orbifolds, but are noninvertible symmetries.  This was discussed in e.g.~\cite{Bhardwaj:2017xup}, which argued for this result at a categorical level, and it is now widely accepted that these symmetries exist.
However,  more concrete and explicit computations verifying these statements are unfortunately not common in the literature.

The purpose of this note is to understand quantum symmetries in examples of two-dimensional nonabelian orbifolds more concretely, by first computing partial trace contributions to partition functions and then explicitly verifying in partition functions that 
gauging the (noninvertible) quantum symmetry returns the original theory.  
This is a followup to one of the authors' work on partition functions in gauged noninvertible symmetries in two dimensions
\cite{Perez-Lona:2023djo,Perez-Lona:2024sds}. See also \cite{Yifan} for related work on noninvertible gaugings, and \cite[\S 4.5]{XY:2024} on Rep$(D_4)$ specifically. 

We begin in section~\ref{sect:rev} by briefly reviewing the historical case of quantum symmetries in abelian orbifolds.
In section~\ref{sect:exs} we study quantum symmetries in several examples of nonabelian orbifolds, with gauge groups $S_3$, $D_4$,
and $Q_8$.  The basic strategy in each case is to begin with a simple ansatz for the insertion of a single quantum symmetry line operator, which determines a subset of the partial traces, and then use modular invariance to determine all other partial traces.  Then, after having constructed a set of partial traces consistent with modular invariance, we check that the partition function of $[ [X/G] / {\rm Rep}(G)]$ equals
that of $X$. We apply this procedure to several examples. For the initial example of Rep$(S_3)$, we walk through the details, then, give a simplified version of that computation, which is repeated for the other examples.  The case of Rep$(S_3)$ was also previously computed using other methods in \cite{Daniel:2024}, and we recover their results.  Our other examples are, to our knowledge, new. 

Here we only deal with multiplicity-free cases.  We believe the same methods apply more generally, but leave more complex examples for future work.

In appendix~\ref{sect:maxwell} we discuss the analogous quantum symmetry computation in two-dimensional pure $U(1)$ Maxwell theory, and in appendix~\ref{app:summ:modtrans} we collect modular transformations of partial traces
computed in \cite{Perez-Lona:2023djo}, to make this paper self-contained.

\section{Review of quantum symmetries in abelian orbifolds}  \label{sect:rev}

Let us quickly review quantum symmetries in two-dimensional ${\mathbb Z}_k$ orbifolds,
and how, after gauging a ${\mathbb Z}_k$ orbifold by its quantum symmetry, one recovers the
original partition function.

Here also, we can try to understand the partial traces $Z_{1,0},\cdots, Z_{k-1,0}$ for ${\rm Rep}(\mathbb{Z}_k)$ by understanding in a similar way that the second index acts as twist fields and the first index acts as time loops in $Z_{a,b}$. The irreps for $\mathbb{Z}_k$ are denoted by $\chi_r$ and $\chi_r(\omega^s)=\omega^{rs},\hspace{0.25cm}r,s\in \{0,1,\cdots k-1\}$, where $\omega=\exp(2\pi i/ k)$.

Following \cite[section 8.5]{Ginsparg}, we begin by considering the action of a single quantum symmetry line on the $T^2$ partition function of the $[X/{\mathbb Z}_k]$ orbifold.  For this case,
\begin{equation}
    Z_{r,1}(\tau)=\frac{1}{k} \sum_{gh=hg}\chi_r(g)Z_{g,h}(\tau).
\end{equation}
Then, modular invariance determines the other partial traces.
One finds
\begin{eqnarray}
Z_{r,s}(\tau) & = &  \frac{1}{k} \sum_{gh=hg}\chi_r(g)\chi_s(h)Z_{g,h}(\tau),
\\
& = &
 \frac{1}{k} \sum_{a,b=0}^{k-1}e^{\frac{2\pi i}{k}(ra+sb)}Z_{a,b}(\tau), \hspace{0.5cm} r,s\in \{0,1,2,...k-1\}.
\end{eqnarray}
where $g=\omega^a, h=\omega^b$.
This coincides with \cite[eq. (2.1)]{Bhardwaj:2017xup} up to $m\leftrightarrow s, n\leftrightarrow -r$ (mod k).

One can quickly confirm that
\begin{eqnarray*}
    Z( [ [X/{\mathbb Z}_k] / {\rm Rep}({\mathbb Z}_k)] ) 
    & = &
    \frac{1}{k} \sum_{r,s} Z_{r,s},
    \\
    & = & 
    Z(X)_{a=0, b=0},
\end{eqnarray*}
using the fact that roots of unity sum to zero.

Thus, we see explicitly $[[X/\mathbb{Z}_k]/{\rm Rep}(\mathbb{Z}_k)]=X$, at least for $T^2$ partition functions.

We will follow the same strategy to determine the partial traces of orbifolds by noninvertible symmetries, namely, start with a simple assumption to determine some of the partial traces, and use modular invariance to determine the rest.  Then, given the results for partial traces determined by modular invariance, in each case we will check that the partition function of the Rep($G$)-gauged $G$-orbifold is the same as the partition function of the original theory.

\section{Examples of non-abelian orbifolds}  \label{sect:exs}

\subsection{Detailed computation in ${\rm Rep}(S_3)$ case}

In this section, we will begin with a simple ansatz for the action of the quantum symmetry, and then apply modular transformations
to derive partial traces, using the modular transformations for general $\beta$'s listed in \cite{Perez-Lona:2023djo}.
Then, after we have derived partial traces, we will require that a Rep$(S_3)$ gauging of an $S_3$ orbifold returns the original theory.
We will see that this requirement plus the form of the original ansatz uniquely determines all remaining unknown coefficients. The results have also been discussed in \cite{Daniel:2024}.

First, trivially, 
\begin{equation}
Z_{1,1}^1(\tau)=\frac{1}{6}\sum_{gh=hg} Z_{g,h}(\tau).
\end{equation}

Now, following the same strategy as \cite{Ginsparg} in the abelian case,
we insert a quantum symmetry line, which corresponds to the partial traces $Z^R_{R,1}$ for any irreducible representation $R$.
We make the initial ansatz that
\begin{equation}
    Z^R_{R,1} \: = \: \frac{1}{6}\sum_{gh=hg}\chi_R(g) Z_{g,h}(\tau)
\end{equation}
for any irreducible representation $R$..
Our strategy is then to apply modular transformations to derive the other partial traces, which we will outline below.

For later use,
\begin{equation*}
\chi_X([1])=1, \quad \chi_X([a])=-1,  \quad \chi_X([b])=1, \quad  \chi_Y([1])=2,  \quad \chi_Y([a])=0, 
\quad \chi_Y([b])=-1.
\end{equation*}

Using the modular transformation, we write
\begin{equation}
    Z_{R,R}^1(\tau)=Z_{R,1}^R(\tau+1),
\end{equation}
which implies
\begin{equation}
Z_{R,R}^1(\tau)=\frac{1}{6} \sum_{gh=hg}\chi_R(g)Z_{gh,h}(\tau)
\end{equation}
for any irreducible representation $R$.

Also, $Z_{1,R}^R(\tau)=Z_{R,1}^R(-1/\tau)$, which implies
\begin{equation}
Z_{1,R}^R(\tau) = \frac{1}{6} \sum_{gh=hg}\chi_R(g)Z_{h^{-1},g}(\tau)
\end{equation}
for any irreducible representation $R$..

Now, doing one more round of $S$, $T$ transformations, we get the following consistency checks:
\begin{eqnarray*}
\sum_{gh=hg}Z_{g,h}  =  \sum_{gh=hg}Z_{gh,h}=\sum_{gh=hg}Z_{h^{-1},g} & \implies & Z_{1,1}^1(\tau+1)=Z_{1,1}^1(\tau)=Z_{1,1}^1(-\frac{1}{\tau}),
\\
\sum_{gh=hg}\chi_X(g)Z_{g,h}  =  \sum_{gh=hg}\chi_X(g)Z_{gh^2,h} & \implies & Z_{X,X}^1(\tau+1)=Z_{X,1}^X(\tau),
\\
\sum_{gh=hg}\chi_X(g)Z_{h^{-1},g}  =  \sum_{gh=hg}\chi_X(g)Z_{h^{-1}g,g} & \implies & Z_{1,X}^X(\tau+1)=Z_{1,X}^X(\tau),
\\
\sum_{gh=hg}\chi_X(g)Z_{g,h}  =  \sum_{gh=hg}\chi_X(g)Z_{g^{-1},h^{-1}} & \implies & Z_{1,X}^X(-\frac{1}{\tau})=Z_{X,1}^X(\tau),
\\
\sum_{gh=hg}\chi_X(g)Z_{gh,h}  =  \sum_{gh=hg}\chi_X(g)Z_{h^{-1},gh} & \implies & Z_{X,X}^1(-\frac{1}{\tau})=Z_{X,X}^1(\tau),
\\
\sum_{gh=hg}\chi_Y(g)Z_{h^{-1},g}  =  \sum_{gh=hg}\chi_Y(g)Z_{h^{-1}g,g} & \implies & Z_{1,Y}^Y(\tau+1)=Z_{1,Y}^Y(\tau),
\\
\sum_{gh=hg}\chi_Y(g)Z_{g,h}  =  \sum_{gh=hg}\chi_Y(g)Z_{g^{-1},h^{-1}} & \implies & Z_{1,Y}^Y(-\frac{1}{\tau})=Z_{Y,1}^Y(\tau).
\end{eqnarray*}

So far, we have passed all the consistency checks \cite[eqns. (3.110-3.114), (3.116), (3.121-3.125), (3.127)]{Perez-Lona:2023djo}. Now, we need to compute $Z_{X,Y}^Y, Z_{Y,X}^Y, Z_{Y,Y}^X, Z_{Y,Y}^Y$.

From \cite[eq.~(3.118)]{Perez-Lona:2023djo},we have
\begin{equation}\label{3.118}\frac{\beta_4}{\beta_3\beta_6} \, Z^Y_{Y,X}(\tau)+\frac{\beta_4}{\beta_5^2} \, Z^Y_{Y,Y}(\tau)=2Z^1_{Y,Y}(\tau+1)-Z^Y_{Y,1}(\tau),\end{equation}
after one more $T$ transform which becomes (using \cite[eqns.~(3.116), (3.117), (3.120)]{Perez-Lona:2023djo})
\begin{equation}\label{3.1182}2Z^1_{Y,Y}(\tau+2)= Z^1_{Y,Y}(\tau)-\frac{\beta_4}{\beta_2\beta_6} \, Z^X_{Y,Y}(\tau)+ Z^Y_{Y,1}(\tau)-\frac{\beta_4}{\beta_3 \beta_6} \, Z^{Y}_{Y,X}(\tau),\end{equation}
which upon addition gives
\begin{equation}\label{3.1182'}2Z^1_{Y,Y}(\tau+2)+2Z^1_{Y,Y}(\tau+1)-Z^1_{Y,Y}(\tau)-2Z^Y_{Y,1}(\tau)=-\frac{\beta_4}{\beta_2\beta_6} Z^X_{Y,Y}(\tau)+\frac{\beta_4}{\beta_5^2}Z^Y_{Y,Y}(\tau).\end{equation}

So far, we have two equations in three unknowns.

Now, we have \cite[eq. (3.129)]{Perez-Lona:2023djo} from $S$ transformation,
\begin{equation}
    \label{3.1292}
  -\frac{\beta_4}{\beta_2\beta_6} \, Z^X_{Y,Y}(\tau)+\frac{\beta_4}{\beta_5^2} \, Z^Y_{Y,Y}(\tau)= 2Z^1_{Y,Y}\left(-\frac{1}{\tau}\right)-Z^1_{Y,Y} (\tau).
\end{equation}
As a consistency check, it can be shown that 
$$2Z^1_{Y,Y}\left(-\frac{1}{\tau}\right) - Z^1_{Y,Y}(\tau)
=
2Z^1_{Y,Y}(\tau+2) + 2Z^1_{Y,Y}(\tau+1) - Z^1_{Y,Y}(\tau) - 2Z^Y_{Y,1}(\tau).$$
This makes the two equations \eqref{3.1182'} and \eqref{3.1292} consistent with each other\footnote{More $T$ transformations generate equivalent conditions, no new information can be obtained from them.}.

Applying $S$ transformations once more on \cite[eqns.~(3.130), (3.131)]{Perez-Lona:2023djo}, we get the same results involving $Z^Y_{Y,Y},Z^X_{Y,Y}$. So, no new information can be obtained from them.

Applying one more $T$ transformation on \eqref{3.1182'} and using \cite[eqns.~(3.119),(3.120)]{Perez-Lona:2023djo} we get
same info as \eqref{3.1182}, so consistent. So again, we do not get any new information from this operation.
So far, everything seems consistent.
Now, we return back to the two independent equations, which we take to be \eqref{3.118}, \eqref{3.1182} in terms of the three unknown partial traces $Z^Y_{Y,Y},Z^X_{Y,Y},Z^Y_{Y,X}$ . We now write down the two equations in terms of the known quantities, involving the trivial irrep. After doing some laborious algebra, and using $$Z^1_{Y,Y}(\tau+k)=\frac{1}{6}\sum_{gh=hg}\chi_Y(g)Z_{gh^k,h}$$ we obtain the results for
\begin{equation}
    \frac{\beta_4}{\beta_5^2} \, Z^Y_{Y,Y}(\tau)+\frac{\beta_4}{\beta_3\beta_6} \, Z^Y_{Y,X}(\tau)
\end{equation}
and 
\begin{equation}
    -\frac{\beta_4}{\beta_2\beta_6}\, Z^X_{Y,Y}(\tau)-\frac{\beta_4}{\beta_3 \beta_6} \, Z^Y_{Y,X}(\tau).
\end{equation}
We now take the most generic form of $Z^Y_{Y,X}$ to be
\begin{equation}
    Z^Y_{Y,X}(\tau)=\frac{1}{6} \sum_{gh=hg} c_{g,h}Z_{g,h}(\tau)
\end{equation}
and plugging this into the last two equations we obtain the results for $\frac{\beta_4}{\beta_5^2} \, Z^Y_{Y,Y}(\tau)$
and 
    $-\frac{\beta_4}{\beta_2 \beta_6} \, Z^X_{Y,Y}(\tau)$ in terms of the $c_{g,h}$'s.
    
Now, we want to apply modular transformations on these two partial traces. Applying our earlier machinery of partial traces under $T$ transformation, we can compute $\frac{\beta_4}{\beta_5^2} \, Z^Y_{Y,Y}(\tau+1)$
and 
    $-\frac{\beta_4}{\beta_2 \beta_6} \, Z^X_{Y,Y}(\tau+1)$ in terms of the $c_{g,h}$'s.
    
From \cite[eqns. (3.120), (3.119)]{Perez-Lona:2023djo} we know these are equal to the following expressions respectively
\begin{eqnarray}
\label{YYY2}
    \frac{\beta_4}{\beta_5^2}\, Z^Y_{Y,Y}(\tau+1) & = &
    Z^Y_{Y,1}(\tau)-\frac{\beta_4}{\beta_3 \beta_6}\, Z^Y_{Y,X}(\tau).
\\
\label{XYY2}
   -\frac{\beta_4}{\beta_2\beta_6}\, Z^X_{Y,Y}(\tau+1) & = & 
   \frac{1}{2} \left(Z^Y_{Y,1}(\tau)+\frac{\beta_4}{\beta_3 \beta_6} \, Z^Y_{Y,X}(\tau)-\frac{\beta_4}{\beta_5^2} \, Z^Y_{Y,Y}(\tau) \right).
\end{eqnarray}
Comparing the coefficients for the two sets, we get various consistency relations between the various coefficients $c_{g,h}$'s.

Now, we would like to check the S transformations as in \cite[eqns. (3.130),
(3.131)]{Perez-Lona:2023djo} and compute
\begin{equation}
    \frac{\beta_4}{\beta_5^2}\, Z^Y_{Y,Y}\left(-\frac{1}{\tau}\right) \mbox{  and  }
    -\frac{\beta_4}{\beta_2 \beta_6}\, Z^X_{Y,Y}\left(-\frac{1}{\tau}\right).
\end{equation}
But from \cite[eqns.~(3.131), (3.130)]{Perez-Lona:2023djo} these are equal to, respectively,
\begin{eqnarray}
\label{s1'}
   \frac{\beta_4}{\beta_5^2}\, Z^Y_{Y,Y}\big(-\frac{1}{\tau}\big) & = &
   Z^1_{Y,Y}(\tau)+\frac{\beta_4}{\beta_2\beta_6} \, Z^X_{Y,Y}(\tau)
   \\
\label{s2'}
        -\frac{\beta_4}{\beta_2\beta_6}\, Z^X_{Y,Y}\big(-\frac{1}{\tau}\big) & = &
        \frac{1}{2} \left(Z^1_{Y,Y}(\tau)-\frac{\beta_4}{\beta_2\beta_6}\, Z^X_{Y,Y}(\tau)-\frac{\beta_4}{\beta_5^2}\, Z^Y_{Y,Y}(\tau) \right).
\end{eqnarray}
Comparing the coefficients we again get another set of relations between the various $c_{g,h}$'s.

Coming back to $Z^Y_{Y,X}(\tau)$ we find \cite[eq.~(3.128)]{Perez-Lona:2023djo}
\begin{equation}
\label{YXY1}
\frac{\beta_1\beta_6}{\beta_2\beta_4} \, Z^Y_{X,Y}(\tau)=Z^Y_{Y,X}\left(-\frac{1}{\tau}\right).
\end{equation}
Applying the $T$ transformation once, we can compute $\frac{\beta_1\beta_6}{\beta_2\beta_4}Z^Y_{X,Y}(\tau+1)$.

Now, we know $Z^Y_{X,Y}(\tau+1)=Z^Y_{X,Y}(\tau)$ from \cite[eq. (3.115)]{Perez-Lona:2023djo}. So comparing the coefficients,we obtain in a similar fashion various relations between the $c_{g,h}$'s.

Applying one S transformation on $Z^Y_{X,Y}$ we have from \cite[eq.~(3.126)]{Perez-Lona:2023djo}
\begin{equation}
\label{YXY3}
\frac{\beta_1\beta_6}{\beta_2\beta_4} \, Z^Y_{X,Y} \left(-\frac{1}{\tau}\right)
=
Z^Y_{Y,X}(\tau).
\end{equation}
Comparing the coefficients we get another set relating the various coefficients.

We have verified all the consistency checks in \cite[eqns. (3.110)-(3.131)]{Perez-Lona:2023djo}.

Thus, from all the previous set of relations amongst the $c_{g,h}$'s we can see that the relations are consistent amongst each other and we have just five independent unknowns for the partial traces
\begin{eqnarray}
       \frac{\beta_4}{\beta_3\beta_6} c_{1,1} & := & s_1,
       \\
        \frac{\beta_4}{\beta_3\beta_6}c_{a,a} & = & \frac{\beta_4}{\beta_3\beta_6}c_{a,1}
         =  \frac{\beta_4}{\beta_3\beta_6}c_{1,a}-2:=s_2,
        \\
        \frac{\beta_4}{\beta_3\beta_6}c_{ab,ab} & = & 
        \frac{\beta_4}{\beta_3\beta_6}c_{ab,1}  =  \frac{\beta_4}{\beta_3\beta_6}c_{1,ab}-2:=s_3,
        \\
        \frac{\beta_4}{\beta_3\beta_6}c_{ab^2,ab^2} & = & 
        \frac{\beta_4}{\beta_3\beta_6}c_{ab^2,1}=\frac{\beta_4}{\beta_3\beta_6}c_{1,ab^2}-2:=s_4,
        \\
   \frac{\beta_4}{\beta_3\beta_6}c_{b,1}
   & = &
   \frac{\beta_4}{\beta_3\beta_6}c_{b,b}=\frac{\beta_4}{\beta_3\beta_6}c_{b,b^2}=\frac{\beta_4}{\beta_3\beta_6}c_{b^2,1}=\frac{\beta_1}{\beta_2\beta_3}c_{b^2,b}
   \\
   & = &
   \frac{\beta_1}{\beta_2\beta_3}c_{b^2,b^2} = \frac{\beta_4}{\beta_3\beta_6}c_{1,b}+3=\frac{\beta_4}{\beta_3\beta_6}c_{1,b^2}+3:=s_5.
\end{eqnarray}
This exhausts all the modular transformations in \cite[eqs.~(3.110)-(3.131)]{Perez-Lona:2023djo}.

To summarize, modular invariance implies that the partial traces are
\begin{eqnarray}
Z^R_{R,1}(\tau) & = & \frac{1}{6} \sum_{gh=hg}\chi_R(g) Z_{g,h},
\\
Z^R_{1,R}(\tau) & = & \frac{1}{6} \sum_{gh=hg}\chi_R(g) Z_{h^{-1},g}=\frac{1}{6} \sum_{gh=hg}\chi_R(h) Z_{g,h},
\\
Z^1_{R,R}(\tau) & = & \frac{1}{6} \sum_{gh=hg}\chi_R(g) Z_{gh,h},
\\
Z^Y_{Y,X}(\tau) & = & \frac{1}{6} \sum_{gh=hg}c_{g,h} Z_{g,h},
\\
\frac{\beta_1\beta_6}{\beta_2\beta_4}Z^Y_{X,Y}(\tau) & = & Z^Y_{Y,X}(-1/\tau)=\frac{1}{6} \sum_{gh=hg}c_{g,h}Z_{h^{-1},g},
\\
Z^Y_{Y,Y}(\tau) & = & \frac{\beta_5^2}{\beta_4} Z^Y_{Y,1}(\tau-1)- \frac{\beta_5^2}{\beta_3 \beta_6}Z^Y_{Y,X}(\tau-1),
\\
& = & \frac{1}{6} \big( \frac{\beta_5^2}{\beta_4} \chi_Y(g) - \frac{\beta_5^2}{\beta_3 \beta_6} c_{g,h}\big) Z_{gh^{-1},h},
\\
  Z^X_{Y,Y}(\tau) & = & -\frac{\beta_2 \beta_6}{2\beta_4}Z^Y_{Y,1}(\tau-1)-\frac{\beta_2}{2\beta_3}Z^Y_{Y,X}(\tau-1)+\frac{\beta_2 \beta_6}{2\beta_5^2} Z^Y_{Y,Y}(\tau-1),
  \nonumber
  \\
   & = & -\frac{\beta_2 \beta_6}{12\beta_4}\chi_Y(g) \big(Z_{gh^{-1},h}- Z_{gh^{-2},h}\big) -\frac{\beta_2}{12\beta_3}c_{g,h} \big(Z_{gh^{-1},h}+Z_{gh^{-2},h}\big).
   \nonumber
 \end{eqnarray}

Next, we check that the partition function of the Rep$(S_3)$ gauged $S_3$ orbifold should return the original theory, before any orbifolding.  We will see that this imposes a constraint on some of the unknowns above.

Recall that the fully ${\rm Rep}(S_3)$ gauged partition function is \cite[eq. (3.211)]{Perez-Lona:2023djo}
\begin{eqnarray}
    Z_{1+X+2Y}= \frac{1}{6} \Big(Z^1_{1,1}+Z^X_{1,X}+Z^1_{X,X}+Z^X_{X,1}+2 Z^Y_{1,Y}+2 Z^Y_{Y,1}+2 Z^1_{Y,Y}
    \nonumber \\
    - 2\frac{\beta_1}{\beta_2\beta_3}  Z^Y_{X,Y}-2 \frac{\beta_4}{\beta_3\beta_6}Z^{Y}_{Y,X} +2\frac{\beta_4}{\beta_2 \beta_6} Z^X_{Y,Y}+2 \frac{\beta_4}{\beta_5^2} Z^Y_{Y,Y}\Big).
    \end{eqnarray}

Now, if we demand that the gauged partition function should give us back $ Z_{1,1}$, then all other contributions need to vanish. After a computation, we see this implies that the $s$'s are given by 
\begin{equation}
    s_1=-2,\hspace{0.5cm}s_2=s_3=s_4=0,\hspace{0.5cm}s_5=1.
\end{equation}
Thus, we have completed our calculation the partial traces of ${\rm Rep}(S_3)$ in terms of the partial traces of Vec$(S_3)$.

\subsection{Simplified ${\rm Rep}(S_3)$ computation}

In this section, we will give a simplified version of the computation of the
previous subsection, and again focus on the partial traces of ${\rm Rep}(S_3)$, where $S_3$ can be presented as \cite[eq. (3.1)]{Perez-Lona:2023djo}
    \begin{equation}
        \langle a,b \, \arrowvert \, a^2=b^3=1, \: a b a = b^2 \rangle.
    \end{equation}
    We have the conjugacy classes 
    \begin{equation}
    [1]=\{1\}, [a]=\{a,ab,ab^2\}, [b]=\{b,b^2\}.
    \end{equation}
    The group has three irreps labelled by $1, X, Y$, where $Y$ is the two dimensional irrep, the others are all one dimensional irreps.  Their characters are listed in table~\ref{table-s3-char}.
   
\begin{table}
    \begin{center}
        \begin{tabular}{c|rrr} 
        & $[1]$ & $[a]$ & $[b]$ \\ \hline
        $1$ & $1$ & $1$ & $1$ \\
        $X$ & $1$ & $-1$ & $1$ \\
        $Y$ & $2$ & $0$ & $-1$
        \end{tabular}
        \caption{Character table for $S_3$. \label{table-s3-char}}
    \end{center}
\end{table}

    The fusion rules are given by \cite[eq.
    (3.5)]{Perez-Lona:2023djo}
    \begin{equation}
        \begin{array}{l}
        X\otimes X\cong 1,\hspace{0.25cm}
        X\otimes Y\cong Y,\hspace{0.25cm}
        Y \otimes Y \cong 1\oplus X \oplus Y.
        \end{array}
    \end{equation}
    We list here the ordered commuting pairs $(g,h)$
    \begin{equation}
        \begin{array}{l}
        g=1, h\in \{1,a,ab,ab^2,b,b^2\},\\
        g=a, h\in\{1,a\},\\
        g=ab, h\in \{1,ab\},\\
         g=ab^2, h\in\{1,ab^2\},\\
        g=b, h\in \{1,b,b^2\},\\
        g=b^2, h\in \{1,b,b^2\}.\\
            \end{array}
        \end{equation}

We proceed much as in \cite[section 8.5]{Ginsparg}.  
The one-dimensional representations form an abelian group, so for partial traces associated to one-dimensional representations, we begin by assuming that the quantum symmetry phase factors are the same as
for the case of abelian groups, namely:
\begin{equation}
    Z^{p \otimes q}_{p,q}(\tau)=\frac{1}{6} \sum_{gh=hg}\chi_p(g)\chi_q(h)Z_{g,h}(\tau).
\end{equation}
where $p$, $q$ are one-dimensional representations (namely, $1$ or $X$).

Then, requiring modular invariance determines the other partial traces.
(See appendix~\ref{app:s3-mod-trans} for the modular transformations.) 
As the algebra required to apply modular transformations is laborious,
we merely list the results for partial traces
below:
\begin{eqnarray}
    Z^{p \otimes q}_{p,q}(\tau) & = & \frac{1}{6} \sum_{gh=hg}\chi_p(g)\chi_q(h) \, Z_{g,h}(\tau),
    \\
    Z^Y_{p,Y}(\tau) & = & \frac{k_p}{6} \sum_{gh=hg}\chi_p(g)\chi_Y(h) \, Z_{g,h}(\tau),
    \\
     Z^Y_{Y,p}(\tau) & = & \frac{k_p}{6} \sum_{gh=hg}\chi_Y(g)\chi_p(h) \, Z_{g,h}(\tau),
     \\
     Z^p_{Y,Y}(\tau) & = & \frac{k_p}{6} \sum_{gh=hg}\chi_Y(g)\chi_p(h) \, Z_{gh,h}(\tau),
     \\
     Z^Y_{Y,Y}(\tau) 
     & = & \frac{1}{6} \sum_{gh=hg} \chi_Y(g) \left( 1 + \chi_X(h) \right) Z_{gh^{-1},h}(\tau),
\end{eqnarray}
where $p \in \{1, X\}$, $k_1=1,\, k_X=-1$.
In particular, it is straightforward to check that the expressions above are consistent with modular invariance, in the sense that the known modular transformation properties of ordinary orbifold partial traces reproduce the modular transformations of the Rep$(S_3)$ partial traces listed in appendix~\ref{app:summ:modtrans}, after the identifications above.
These match with the results in \cite[eqns. (3.61)-(3.71)]{Daniel:2024} upto some other conventions.\footnote{In \cite{Daniel:2024}, they have taken that under T transformation $Z_{g,h}\rightarrow Z_{g,hg^{-1}}$ and here, we are taking $T:Z_{g,h}\rightarrow Z_{gh,h}$. Moreover, the character products with $g,h$ swapped in our set of formulae would give us their set of results. Say, we take $Z^{p \otimes q}_{p,q}(\tau)= \frac{1}{6} \sum_{gh=hg}\chi_p(h)\chi_q(g) \, Z_{g,h}(\tau)$ and so on for the other partial traces according to their results. Lastly, they have combined elements of the same conjugacy class as one element with factors of the size of the conjugacy class appearing before them.}

In this fashion, we derive
all the (genus one) partial traces of ${\rm Rep}(S_3)$ using modular transformations, from an initial assumption that for the subsymmetry defined by one-dimensional representations, the quantum symmetry action is the same as for an abelian theory.

Now that we have determined the partial traces, let us check that, at the level of partition functions,
the Rep$(S_3)$ symmetry is indeed acting as a quantum symmetry.
Upon plugging everything in the gauged partition function \cite[ eq.~(3.617)]{Perez-Lona:2023djo} 
\begin{displaymath}
Z_{1+X+2Y} = \frac{1}{6} \Big(Z^1_{1,1}+Z^X_{1,X}+Z^X_{X,1}+Z^1_{X,X}+2 Z^Y_{1,Y}+2Z^Y_{Y,1}+2Z^1_{Y,Y}-2Z^Y_{X,Y}-2Z^Y_{Y,X}-2Z^X_{Y,Y}+2Z^Y_{Y,Y}\Big)
\end{displaymath}
we get the desired result, 
\begin{equation}
    Z( [ [X/S_3] / {\rm Rep}(S_3)] ) \: = \: Z_{1,1} \: = \: Z(X).
\end{equation}

Our initial assumption, that the partial traces for one-dimensional representations match those of an abelian group, has been validated by the existence of an extension to a complete modular-invariant set of partial traces, such that after gauging the Rep $S_3$ quantum symmetry of an $S_3$ orbifold, we recover the original theory.  

In the next several subsections we will apply the same strategy to other examples.
In each case, we begin with the initial assumption that for one-dimensional representations, the quantum symmetry acts in the same way as for an abelian orbifold.  We then use modular transformations to determine a full modular-invariant set of partial traces for a Rep$(G)$ orbifold.  Finally, given the results determined by modular invariance, we check in each case that orbifolding a $G$-orbifold by the Rep$(G)$ quantum symmetry returns the partition function of the original theory, as expected.

\subsection{${\rm Rep}(D_4)$}
    In this section, we will focus on the partial traces of ${\rm Rep}(D_4)$, where $D_4$ can be presented as \cite[eq. (3.217)]{Perez-Lona:2023djo}
    \begin{equation}
        \langle x,y \, \arrowvert \, x^4 = y^2 = (xy)^2 = 1\rangle,
    \end{equation}
    with conjugacy classes 
    \begin{equation}
    [1]=\{1\}, [x]=\{x,x^3\}, [x^2]=\{x^2\}, [y]=\{y,x^2y\}, [xy]=\{xy, x^3y\}.
    \end{equation}
    The group has five irreducible representations labelled by $1, a, b, c, m$.  Of these, $m$ is two-dimensional, the others are one-dimensional, and their characters are listed in table~\ref{table-d4-char}.

\begin{table}
    \begin{center}
        \begin{tabular}{c|rrrrr}
        & $[1]$ & $[x^2]$ & $[x]$ & $[y]$ & $[xy]$ \\ \hline
        $1$ & $1$ & $1$ & $1$ & $1$ & $1$ \\
        $a$ & $1$ & $1$ & $1$ & $-1$ & $-1$ \\
        $b$ & $1$ & $1$ & $-1$ & $1$ & $-1$ \\
        $c$ & $1$ & $1$ & $-1$ & $-1$ & $1$ \\
        $m$ & $2$ & $-2$ & $0$ & $0$ & $0$
    \end{tabular}
    \caption{Character table for $D_4$.  \label{table-d4-char} }
    \end{center}
\end{table}

The fusion rules are given by \cite[eq.~(3.218)]{Perez-Lona:2023djo}
    \begin{equation}
        \begin{array}{l}
        a\otimes a\cong b\otimes b\cong c\otimes c\cong 1,\\
        a\otimes b\cong c, \hspace{0.25cm} a\otimes c\cong b,\hspace{0.25cm} b\otimes c\cong a,\\
        a\otimes m\cong b\otimes m \cong c\otimes m\cong m,\\
        m\otimes m\cong 1\oplus a\oplus b\oplus c.
        \end{array}
    \end{equation}
    We list here the ordered commuting pairs $(g,h)$
    \begin{equation}
        \begin{array}{l}
        g=1, h\in \{1,x,x^2,x^3,y,xy,x^2y,x^3y\},\\
        g=x, h\in\{1,x,x^2,x^3\},\\
        g=x^2, h\in \{1,x,x^2,x^3,y,xy,x^2y,x^3y\},\\
         g=x^3, h\in\{1,x,x^2,x^3\},\\
        g=y, h\in \{1,x^2,y,x^2y\},\\
        g=xy, h\in \{1,x^2,xy,x^3y\},\\
        g=x^2y, h\in \{1,x^2,y,x^2y\},\\
        g=x^3y, h\in \{1,x^2,xy,x^3y\}.
            \end{array}
\end{equation}

We proceed much like \cite[section 8.5]{Ginsparg}.  
The one-dimensional representations form a group, so for partial traces associated to one-dimensional representations, we begin by assuming that the quantum symmetry phase factors are the same as
for the case of abelian groups, namely:
\begin{equation}
    Z^{p \otimes q}_{p,q}(\tau)=\frac{1}{8} \sum_{gh=hg}\chi_p(g)\chi_q(h)Z_{g,h}(\tau).
\end{equation}
Then, requiring modular invariance determines the other partial traces.
(See appendix~\ref{app:d4-mod-trans} for the modular transformations.)
We summarize the results below:
\begin{eqnarray}
    Z^{p \otimes q}_{p,q}(\tau) & = & \frac{1}{8} \sum_{gh=hg} \chi_p(g)\chi_q(h) \, Z_{g,h}(\tau),
    \\
    Z^m_{p,m}(\tau) & = & \frac{k_p}{8} \sum_{gh=hg} \chi_p(g)\chi_m(h) \, Z_{g,h}(\tau),
    \\
     Z^m_{m,p}(\tau) & = & \frac{k_p}{8} \sum_{gh=hg} \chi_m(g)\chi_p(h) \, Z_{g,h}(\tau),
     \\
     Z^p_{m,m}(\tau) & = & \frac{k_p}{8} \sum_{gh=hg} \chi_m(g)\chi_p(h) \, Z_{gh,h}(\tau),
\end{eqnarray}
where $k_1=1,\, k_a=-1,\, k_b=1,\, k_c=1$.
In particular, it is straightforward to check that the expressions above determine a modular-invariant partition function.

In this fashion, we derive
all the (genus one) partial traces of ${\rm Rep}(D_4)$ using modular transformations, from an initial assumption that for the subsymmetry defined by one-dimensional representations, the quantum symmetry action is the same as for an abelian theory.

Now that we have determined the partial traces,
let us check that, at the level of partition functions,
the Rep$(D_4)$ symmetry is indeed acting as a quantum symmetry.
Plugging everything into the expression for the Rep$(D_4)$-gauged partition function \cite[eq.~(3.625)]{Perez-Lona:2023djo}
\begin{eqnarray}
Z_{1+a+b+c+2m} & = &
\frac{1}{8} \Big(Z^1_{1,1}+Z^a_{1,a}+Z^a_{a,1}+Z^1_{a,a}+Z^b_{1,b}+Z^b_{b,1}+Z^1_{b,b}+ Z^c_{1,c}+Z^c_{c,1}+Z^1_{c,c}
\nonumber \\
&&
+Z^c_{a,b}+Z^b_{a,c}+Z^c_{b,a}
+Z^a_{b,c}+Z^b_{c,a}+Z^a_{c,b}
\nonumber \\
&& 
+2 Z^m_{1,m}+2Z^m_{m,1}+2 Z^1_{m,m}-2 Z^m_{a,m}-2 Z^m_{m,a}-2 Z^a_{m,m}
\nonumber \\
& & +2 Z^m_{b,m}+2Z^m_{m,b}+2 Z^b_{m,m}+2 Z^m_{c,m}+2Z^m_{m,c}+2 Z^c_{m,m}\Big)
\end{eqnarray}
we get the desired result
\begin{equation}
    Z( [ [X/D_4] / {\rm Rep}(D_4)] ) \: = \: Z_{1,1} \: = \: Z(X).
\end{equation}

\subsection{${\rm Rep}(Q_8)$}

    In this section, we will focus on the partial traces of ${\rm Rep}(Q_8)$, where $Q_8$ can be presented as \cite[eq.~(3.430)]{Perez-Lona:2023djo}
    \begin{equation}
        \langle x,y \, \arrowvert \, x^2 = y^2 = (xy)^2, x^4=1\rangle,
    \end{equation}
    with conjugacy classes 
    \begin{equation}
    [1]=\{1\}, [x]=\{x,x^3\}, [x^2]=\{x^2\}, [y]=\{y,y^3\}, [xy]=\{xy, x^3y\}.
    \end{equation}
    The group has five irreducible represntations labelled by $1, a, b, c, m$.  Of these, $m$ is two-dimensional, the others are all one-dimensional, and their characters are listed in table~\ref{table-q8-char}.

\begin{table}
    \begin{center}
        \begin{tabular}{c|rrrrr}
        & $[1]$ & $[x^2]$ & $[x]$ & $[y]$ & $[xy]$ \\ \hline
        $1$ & $1$ & $1$ & $1$ & $1$ & $1$ \\
        $a$ & $1$ & $1$ & $1$ & $-1$ & $-1$ \\
        $b$ & $1$ & $1$ & $-1$ & $1$ & $-1$ \\
        $c$ & $1$ & $1$ & $-1$ & $-1$ & $1$ \\
        $m$ & $2$ & $-2$ & $0$ & $0$ & $0$ 
        \end{tabular}
        \caption{Character table for $Q_8$.  \label{table-q8-char}}
    \end{center}
\end{table}
    
The fusion rules are given by \cite[eq.~(3.431)]{Perez-Lona:2023djo}
    \begin{equation}
        \begin{array}{l}
        a\otimes a\cong b\otimes b\cong c\otimes c\cong 1,\\
        a\otimes b\cong c, \hspace{0.25cm} a\otimes c\cong b,\hspace{0.25cm} b\otimes c\cong a,\\
        a\otimes m\cong b\otimes m \cong c\otimes m\cong m,\\
        m\otimes m\cong 1\oplus a\oplus b\oplus c.
        \end{array}
    \end{equation}
    We list here the ordered commuting pairs $(g,h)$
    \begin{equation}
        \begin{array}{l}
        g=1, h\in \{1,x,x^2,x^3,y,xy,x^2y,x^3y\},\\
        g=x, h\in\{1,x,x^2,x^3\},\\
        g=x^2, h\in \{1,x,x^2,x^3,y,xy,x^2y,x^3y\},\\
         g=x^3, h\in\{1,x,x^2,x^3\},\\
        g=y, h\in \{1,x^2,y,x^2y\},\\
        g=xy, h\in \{1,x^2,xy,x^3y\},\\
        g=x^2y, h\in \{1,x^2,y,x^2y\},\\
        g=x^3y, h\in \{1,x^2,xy,x^3y\}.
            \end{array}
        \end{equation}

We proceed much like \cite[section 8.5]{Ginsparg}.  
The one-dimensional representations form a group, so for partial traces associated to one-dimensional representations, we begin by assuming that the quantum symmetry phase factors are the same as
for the case of abelian groups, namely:
\begin{equation}
    Z^{p \otimes q}_{p,q}(\tau)=\frac{1}{8} \sum_{gh=hg}\chi_p(g)\chi_q(h)Z_{g,h}(\tau).
\end{equation}
Then, requiring modular invariance determines the other partial traces.
(See appendix~\ref{app:q8-mod-trans} for the modular transformations.) 

We summarize the partial traces below:
\begin{eqnarray}
    Z^{p \otimes q}_{p,q}(\tau) & = & \frac{1}{8} \sum_{gh=hg}\chi_p(g)\chi_q(h) \, Z_{g,h}(\tau),
    \\
    Z^m_{p,m}(\tau) & = & \frac{k_p}{8} \sum_{gh=hg}\chi_p(g)\chi_m(h) \, Z_{g,h}(\tau),
    \\
     Z^m_{m,p}(\tau) & = & \frac{k_p}{8} \sum_{gh=hg}\chi_m(g)\chi_p(h) \, Z_{g,h}(\tau),
     \\
     Z^p_{m,m}(\tau) & = & \frac{k_p}{8} \sum_{gh=hg}\chi_m(g)\chi_p(h) \, Z_{gh,h}(\tau),
\end{eqnarray}
where $k_1=1, \, k_a=-1, \, k_b=-1, \, k_c = -1$. In particular, it is straightforward to check that the expressions above determine a modular-invariant partition function.

In this fashion, we derive
all the (genus one) partial traces of ${\rm Rep}(Q_8)$ using modular transformations, from an initial assumption that for the subsymmetry defined by one-dimensional representations, the quantum symmetry action is the same as for an abelian theory.

Now that we have determined the partial traces, let us check that, at the level of partition functions,
the Rep$(Q_8)$ symmetry is indeed acting as a quantum symmetry.
Plugging everything into
the gauged partition function \cite[eq.~(3.630)]{Perez-Lona:2023djo}
 \begin{eqnarray}
Z_{1+a+b+c+2m} & = &
\frac{1}{8} \Big(Z^1_{1,1}+Z^a_{1,a}+Z^a_{a,1}+Z^1_{a,a}+Z^b_{1,b}+Z^b_{b,1}+Z^1_{b,b}+ Z^c_{1,c}+Z^c_{c,1}+Z^1_{c,c}
\nonumber \\
&&
+Z^c_{a,b}+Z^b_{a,c}+Z^c_{b,a}
+Z^a_{b,c}+Z^b_{c,a}+Z^a_{c,b}
\nonumber \\
&&
+2 Z^m_{1,m}+2Z^m_{m,1}+2 Z^1_{m,m}-2 Z^m_{a,m}-2 Z^m_{m,a}-2 Z^a_{m,m}
\nonumber \\
& &
-2 Z^m_{b,m}-2Z^m_{m,b}-2 Z^b_{m,m}-2 Z^m_{c,m}-2Z^m_{m,c}-2 Z^c_{m,m}\Big)
\end{eqnarray}
we get the desired result
\begin{equation}
    Z( [ [X/Q_8] / {\rm Rep}(Q_8)] ) \: = \: Z_{1,1} \: = \: Z(X).
\end{equation}

\section{Conclusions}

In this paper we have computed the explicit form of noninvertible quantum Rep$(G)$ actions on two-dimensional $G$ orbifolds, at the level of partition functions, for the nonabelian groups $G = S_3, D_4, Q_8$, to make more explicit how noninvertible quantum symmetries operate.

We have only considered multiplicity-free examples in this note.  We expect that the same ideas should apply more generally, but leave that for future work.

\section{Acknowledgements}
   
We would like to thank A.~Perez-Lona, D.~Robbins, S.~Roy, and H.~Zhang for useful discussions.
E.S.~was partially supported by NSF grant PHY-2310588.

\appendix

\section{Quantum symmetries in 2d Maxwell theory}  \label{sect:maxwell}

In principle, quantum symmetries apply to more than just ordinary orbifolds and 
orbifolds by fusion categories.  In principle, the idea also applies to ordinary gauge theories.
In this section, we describe quantum symmetries and partition function computations in two-dimensional Maxwell theories, to make this point.

We start with $U(1)$ partial traces, denoted by $Z({\theta,\phi}):=Z_{e^{i\theta}e^{i\phi}},$ where $\theta,\phi\in [0,2\pi)$.
The $U(1)$ characters are defined by $\chi_n(e^{i\theta})=\exp(in\theta)$.

Define the Fourier transformed partial traces for the ${\rm Rep}(U(1))\cong \mathbb{Z}$ by
\begin{equation}
    \tilde Z_{m,n}=\frac{1}{4\pi^2}\int d\theta d\phi e^{-im\theta}e^{in\phi} Z({\theta,\phi}),\hspace{0.5cm}{m,n}\in\mathbb{Z}.
\end{equation}

The inverse transform reads
\begin{equation}
    Z(\theta,\phi) = \sum_{m,n\in\mathbb{Z}}e^{im\theta}e^{-in\phi}\tilde Z_{m,n}.
\end{equation}

We want to see the modular transformations on these partial traces on both sides.
\begin{equation}
T:Z_{g.h}\rightarrow Z_{gh,h},\hspace{0.5cm}S:Z_{g,h}\rightarrow Z_{h^{-1},g}.
\end{equation}
For abelian groups, this reduces to 
$T:(g,h)\rightarrow (g+h,h),\hspace{0.5cm} S:(g,h)\rightarrow (-h,g)$.

Thus, for $U(1)$, we have
\begin{equation}
T:Z(\theta,\phi)\rightarrow Z(\theta+\phi,\phi),\hspace{0.5cm}S:Z(\theta,\phi)\rightarrow Z(-\phi,\theta).
\end{equation}

Now, coming to the $\mathbb{Z}$ partial traces
\begin{equation}
    T:\tilde Z_{m,n}\rightarrow \frac{1}{4\pi^2}\int d\theta d\phi e^{-im\theta}e^{in\phi}Z(\theta+\phi,\phi).
\end{equation}
Relabeling angular variables as $\theta'=\theta+\phi$ we see $e^{-im\theta}e^{in\phi}=e^{-im\theta'}e^{i(n+m)\phi}$
which results in 
$$T:\tilde Z_{m,n}\rightarrow \tilde Z_{m,n+m}.$$
Coming to the $S$ transformation, we see
\begin{equation}
S:\tilde Z_{m,n}\rightarrow \frac{1}{4\pi^2}\int d\theta d\phi e^{-im\theta} e^{in\phi} Z(-\phi,\theta).
\end{equation}
Relabeling indices as $\theta'=-\phi,\phi'=\theta$, we see
\begin{equation}
    S:\tilde Z_{m,n}\rightarrow \tilde Z_{n,-m}.
\end{equation}

Now, coming to the double gauging partition function we see
\begin{equation}
    Z_{\mathrm{gauged}} = \sum_{m,n\in \mathbb{Z}}\tilde Z_{m,n}
    = 
    \frac{1}{4\pi^2}\int d\theta d\phi \left(\sum_{m\in\mathbb{Z}} e^{-im\theta} \right)\left(\sum_{n\in\mathbb{Z}}e^{in\phi}\right) Z(\theta,\phi).
\end{equation}
We use the periodic delta-comb result, 
\begin{equation}
    \sum_{m\in\mathbb{Z}}e^{-im\theta}=\sum_{k\in \mathbb{Z}}\delta(\theta-2\pi k), 
    \hspace{0.5cm} \sum_{n\in\mathbb{Z}}e^{in\phi}=\sum_{l\in \mathbb{Z}}\delta(\phi-2\pi l). 
\end{equation}
Here we choose the fundamental domain $\theta,\phi\in [0,2\pi)$ and thus set $k=l=0.$

Then we see that $Z_{\mathrm{gauged}}=Z(0,0)=Z_{1,1}$ as expected.

To understand the partial traces more physically in 2d pure Maxwell, the partition function with two holonomy insertions $U_1,U_2$ with spatial circle circumference L and Euclidean time circumference T is given by \cite[eq. (29)]{Paniak:2003xm}
\begin{equation}
    Z(T,L;U_1,U_2)=\langle U_1 \vert e^{-TH}\vert U_2\rangle=\sum_R \chi_R(U_1)\chi_R(U_2^{\dagger}) e^{-\frac{TL n^2}{2}}
\end{equation}
where $U_1=e^{i\theta},U_2=e^{i\phi}$ giving 
\begin{equation}
    Z(T,L;\theta,\phi)=\sum_{n\in \mathbb{Z}}e^{in(\theta-\phi)}e^{-\frac{An^2}{2}},
\end{equation}
where $A=LT$ is the area.
Here, $\langle \theta \vert n\rangle=\exp(in\theta), \langle n \vert \phi \rangle=\exp(-in\phi)$.

In the dual $\mathbb{Z}$ basis, we have the partial traces to be 
\begin{equation}
\tilde{Z}_{m,n}=\langle m \vert e^{-TH}\vert n\rangle= \delta_{m,n} e^{-\frac{An^2}{2}}
\end{equation}
thus giving 
\begin{equation}
    Z(\theta,\phi) = \sum_{m,n \in \mathbb{Z}} e^{im\theta}e^{-in\phi}\tilde{Z}_{m,n}.
\end{equation}

\section{Modular transformations}  \label{app:summ:modtrans}

In this appendix we collect the modular transformations listed in \cite{Perez-Lona:2023djo}, to make this paper self-contained. The values of the $\beta$'s in \cite{Perez-Lona:2023djo} can be obtained by consistent modular transformations listed in this section. 

\subsection{${\rm Rep}(S_3)$} \label{app:s3-mod-trans}

We  take the values of the $\beta$'s in \cite[eq. (3.612)]{Perez-Lona:2023djo} 
\begin{equation}
    \beta_2=\beta_4=\beta_5=1, 
    \hspace{0.25cm}\beta_1=\beta_3=\beta_6=-1.
\end{equation}
Here, we list the modular transformation equations for easy reference \cite[eqns. (3.110-3.131), (3.612)]{Perez-Lona:2023djo}:
\begin{equation}
    Z^{p\otimes q}_{p,q}(\tau+1)  =  Z^q_{p, p\otimes q}(\tau),
    \quad Z^Y_{p,Y}(\tau+1)=Z^Y_{p,Y}(\tau),
    \quad Z^Y_{Y,p}(\tau+1)= Z^p_{Y,Y}(\tau), 
\end{equation}
\begin{eqnarray}
Z^1_{Y,Y}(\tau+1) & = & \frac{1}{2} (Z^Y_{Y,1}+Z^Y_{Y,X}+Z^Y_{Y,Y})(\tau),\\
Z^X_{Y,Y}(\tau+1) & = & \frac{1}{2} (Z^Y_{Y,1}+Z^Y_{Y,X}-Z^Y_{Y,Y})(\tau),\\
Z^Y_{Y,Y}(\tau+1) & = & Z^Y_{Y,1}(\tau)-Z^Y_{Y,X}(\tau),
\end{eqnarray}
\begin{equation}
Z^{p\otimes q}_{p,q}(-1/\tau)=Z^{p\otimes q}_{q,p}(\tau),
\quad Z^Y_{p,Y}(-1/\tau)=Z^Y_{Y,p}(\tau),
\quad Z^Y_{Y,p}(-1/\tau)=Z^Y_{p,Y}(\tau),
\end{equation}
\begin{eqnarray}
Z^1_{Y,Y}(-1/\tau) & = & \frac{1}{2} (Z^1_{Y,Y}+Z^X_{Y,Y}+Z^Y_{Y,Y})(\tau),\\
Z^X_{Y,Y}(-1/\tau) & = & \frac{1}{2} (Z^1_{Y,Y}+Z^X_{Y,Y}-Z^Y_{Y,Y})(\tau),\\
Z^Y_{Y,Y}(-1/\tau) & = & Z^1_{Y,Y}(\tau)-Z^X_{Y,Y}(\tau),
\end{eqnarray}
where $p,q \in \{1, X\}$

\subsection{${\rm Rep}(D_4)$}   \label{app:d4-mod-trans}

We take the values of $\beta$'s to be \cite[eqns. (3.618)]{Perez-Lona:2023djo}
 \begin{equation}
 \beta_1=\beta_4^2, \quad \beta_3=\frac{\beta_4^2}{\beta_2}, \quad \beta_5=-\beta_2, \quad \beta_6=\beta_4^2, \quad \beta_7=-\frac{\beta_4^2}{\beta_2}, 
 \quad \beta_8=\pm \beta_4, \quad \beta_9=-\frac{\beta_4^2}{\beta_2},
 \end{equation}
 \begin{equation}
 \beta_{10}=\frac{\beta_4^2}{\beta_2}, \quad \beta_{11}=-\frac{\beta_4^2}{\beta_2}, 
 \quad \beta_{12}=\mp \frac{\beta_4^2}{\beta_2}, \quad \beta_{13}=-\beta_4, \quad \beta_{14}=\pm \beta_4, \quad \beta_{15}=\mp\frac{\beta_4^2}{\beta_2},
 \end{equation}
 \begin{equation}
 \beta_{17}=\frac{\beta_2 \beta_{16}}{\beta_4}, \quad \beta_{18}=\pm \beta_4\beta_{16}, \quad \beta_{19}=\pm \beta_{16}.
 \end{equation}

The expressions above reduce all of the $\beta$'s to functions of just $\beta_2$, $\beta_4$, and $\beta_{16}$.  However,
in the modular transformations, those particular $\beta$'s cancel out, so that the choices above suffice to remove all $\beta$-dependence from the
modular transformations.

We list the modular transformations here \cite[eqns. (3.318)-(3.367), (3.618)]{Perez-Lona:2023djo}:
\begin{equation}
    Z^{p\otimes q}_{p,q}(\tau+1)=Z^q_{p, p\otimes q}(\tau),\quad Z^m_{p,m}(\tau+1)=Z^m_{p,m}(\tau),\quad Z^m_{m,p}(\tau+1)= Z^p_{m,m}(\tau),
\end{equation}
\begin{eqnarray}
   Z^1_{m,m}(\tau+1) & = & \frac{1}{2} \Big(Z^m_{m,1}+Z^m_{m,a}+Z^m_{m,b}+Z^m_{m,c}\Big)(\tau),\\
     Z^a_{m,m}(\tau+1) & = & \frac{1}{2} \Big(Z^m_{m,1}+Z^m_{m,a}-Z^m_{m,b}-Z^m_{m,c}\Big)(\tau),\\
      Z^b_{m,m}(\tau+1) & = & \frac{1}{2} \Big(Z^m_{m,1}-Z^m_{m,a}+Z^m_{m,b}-Z^m_{m,c}\Big)(\tau),\\
       Z^c_{m,m}(\tau+1) & = & \frac{1}{2} \Big(Z^m_{m,1}-Z^m_{m,a}-Z^m_{m,b}+Z^m_{m,c}\Big)(\tau),
\end{eqnarray}
\begin{equation}
       Z^{p\otimes q}_{p,q}(-1/\tau)=Z^{p\otimes q}_{q,p}(\tau),\quad Z^m_{p,m}(-1/\tau)=Z^m_{m,p}(\tau),\quad Z^m_{m,p}(-1/\tau)=Z^m_{p,m}(\tau),
\end{equation}
\begin{eqnarray}
         Z^1_{m,m}(-1/\tau) & = & \frac{1}{2} \Big(Z^1_{m,m}+Z^a_{m,m}+Z^b_{m,m}+Z^c_{m,m}\Big)(\tau),\\
          Z^a_{m,m}(-1/\tau) & = & \frac{1}{2} \Big(Z^1_{m,m}+Z^a_{m,m}-Z^b_{m,m}-Z^c_{m,m}\Big)(\tau),\\
           Z^b_{m,m}(-1/\tau) & = & \frac{1}{2} \Big(Z^1_{m,m}-Z^a_{m,m}+Z^b_{m,m}-Z^c_{m,m}\Big)(\tau),\\
            Z^c_{m,m}(-1/\tau) & = & \frac{1}{2} \Big(Z^1_{m,m}-Z^a_{m,m}-Z^b_{m,m}+Z^c_{m,m}\Big)(\tau),
\end{eqnarray}
where $p, q\in \{1, a, b, c\}$.

\subsection{${\rm Rep}(Q_8)$} \label{app:q8-mod-trans}

We take the values of $\beta$'s to be \cite[eqns. (3.626)]{Perez-Lona:2023djo}\footnote{We are writing $\beta$ here instead of $\beta'$.}
\begin{equation}
 \beta_1=\beta_4^2,\quad \beta_3=\frac{\beta_4^2}{\beta_2},\quad \beta_5=-\beta_2,\quad \beta_6=\beta_4^2,\quad \beta_7=-\frac{\beta_4^2}{\beta_2},\quad \beta_8=\pm i\beta_4,\quad \beta_9=-\frac{\beta_4^2}{\beta_2},
 \end{equation}
 \begin{equation}
 \beta_{10}=\frac{\beta_4^2}{\beta_2}, \quad \beta_{11}=-\frac{\beta_4^2}{\beta_2}, \quad \beta_{12}=\mp i\frac{\beta_4^2}{\beta_2}, \quad \beta_{13}=-\beta_4, \quad \beta_{14}=\mp i \beta_4, \quad \beta_{15}=\pm i\frac{\beta_4^2}{\beta_2},
 \end{equation}
 \begin{equation}
 \beta_{17}=\frac{\beta_2 \beta_{16}}{\beta_4}, \quad \beta_{18}=\pm i \beta_4\beta_{16}, \quad \beta_{19}=\pm i\beta_{16}.
 \end{equation}

The expressions above reduce all of the $\beta$'s to functions of just $\beta_2$, $\beta_4$, and $\beta_{16}$.  However,
in the modular transformations, those particular $\beta$'s cancel out, so that the choices above suffice to remove all $\beta$-dependence from the
modular transformations.
 
      We list the modular transformations here \cite[eqns.~(3.483)-(3.496), (3.626)]{Perez-Lona:2023djo}
\begin{equation}
    Z^{p\otimes q}_{p,q}(\tau+1)=Z^q_{p, p\otimes q}(\tau),
    \quad Z^m_{p,m}(\tau+1)=Z^m_{p,m}(\tau),
    \quad Z^m_{m,p}(\tau+1)= Z^p_{m,m}(\tau),  
\end{equation}
\begin{eqnarray}
      Z^1_{m,m}(\tau+1) & = & -\frac{1}{2} \Big(Z^m_{m,1}+Z^m_{m,a}+Z^m_{m,b}+Z^m_{m,c})\Big)(\tau), \\
       Z^a_{m,m}(\tau+1) & = & \frac{1}{2} \Big(-Z^m_{m,1}-Z^m_{m,a}+Z^m_{m,b}+Z^m_{m,c})\Big)(\tau), \\
        Z^b_{m,m}(\tau+1) & = & \frac{1}{2} \Big(-Z^m_{m,1}+Z^m_{m,a}-Z^m_{m,b}+Z^m_{m,c})\Big)(\tau), \\
         Z^c_{m,m}(\tau+1) & = & \frac{1}{2} \Big(-Z^m_{m,1}+Z^m_{m,a}+Z^m_{m,b}-Z^m_{m,c})\Big)(\tau), 
\end{eqnarray}
\begin{equation}      
         Z^{p\otimes q}_{p,q}(-1/\tau)=Z^{p\otimes q}_{q,p}(\tau),
         \quad Z^m_{p,m}(-1/\tau)=Z^m_{m,p}(\tau),
         \quad Z^m_{m,p}(-1/\tau)=Z^m_{p,m}(\tau),
\end{equation}
\begin{eqnarray}
         Z^1_{m,m}(-1/\tau) & = & -\frac{1}{2} \Big(Z^m_{m,1}+Z^m_{m,a}+Z^m_{m,b}+Z^m_{m,c})\Big)(\tau), \\
         Z^a_{m,m}(-1/\tau) & = & \frac{1}{2} \Big(-Z^m_{m,1}-Z^m_{m,a}+Z^m_{m,b}+Z^m_{m,c})\Big)(\tau), \\
        Z^b_{m,m}(-1/\tau) & = & \frac{1}{2} \Big(-Z^m_{m,1}+Z^m_{m,a}-Z^m_{m,b}+Z^m_{m,c})\Big)(\tau), \\
         Z^c_{m,m}(-1/\tau) & = & \frac{1}{2} \Big(-Z^m_{m,1}+Z^m_{m,a}+Z^m_{m,b}-Z^m_{m,c})\Big)(\tau),
\end{eqnarray}
where $p, q\in \{1, a, b, c\}$.

\end{document}